\documentclass[aps,prl,twocolumn,superscriptaddress]{revtex4-2}

\usepackage{amsmath}
\usepackage{amsfonts}
\usepackage{amssymb}
\usepackage{dcolumn}
\usepackage[english]{babel}
\usepackage{epsfig}
\usepackage{natbib}
\usepackage{graphics}
\usepackage{xcolor}
\newcommand{\abs}[1]{\left| #1 \right|}

\begin{document}

\title{Optimal Diffractive Focusing of Quantum Waves}

\author{Maxim A. Efremov}
\affiliation{German Aerospace Center (DLR), Institute of Quantum Technologies, 89081 Ulm, Germany}
\affiliation{Institut f\"ur Quantenphysik and Center for Integrated Quantum Science and Technology ($\it IQST$), Universit\"at Ulm, 89081 Ulm, Germany}

\author{Felix Hufnagel}
\affiliation{Nexus for Quantum Technologies, University of Ottawa, K1N 5N6, Ottawa, ON, Canada}

\author{Hugo Larocque}
\affiliation{Nexus for Quantum Technologies, University of Ottawa, K1N 5N6, Ottawa, ON, Canada}
\affiliation{Research Laboratory of Electronics, Massachusetts Institute of Technology, Cambridge, Massachusetts 02139, USA}

\author{Wolfgang P. Schleich}
\affiliation{Institut f\"ur Quantenphysik and Center for Integrated Quantum Science and Technology ($\it IQST$), Universit\"at Ulm, 89081 Ulm, Germany}
\affiliation{Hagler Institute for Advanced Study at Texas A$\&$M University, Texas A$\&$M AgriLife Research, Institute for Quantum Science and Engineering (IQSE), and Department of Physics and Astronomy, Texas A$\&$M University, College Station, Texas 77843-4242, USA}

\author{Ebrahim Karimi}
\affiliation{Nexus for Quantum Technologies, University of Ottawa, K1N 5N6, Ottawa, ON, Canada}

\date{\today}

\begin{abstract}
Following the familiar analogy between the optical paraxial wave equation and the Schr\"odinger equation, we derive the optimal, real-valued wave function for focusing in one and two space dimensions without the use of any phase component. We compare and contrast the focusing parameters of the optimal waves with those of other diffractive focusing approaches, such as Fresnel zones. Moreover, we experimentally demonstrate these focusing properties on optical beams using both reflective and transmissive liquid crystal devices. Our results provide an alternative direction for focusing waves where phase elements are challenging to implement, such as for X-rays, THz radiation, and electron beams.
\end{abstract}


\maketitle

\noindent\textbf{Introduction}\\ 
Fresnel zone plates \cite{born2013principles} are optical elements that focus an incident beam due to binary variations in its amplitude and phase. They offer precise control over diffractive propagation and enable efficient beam focusing in systems, where traditional lensing elements are not immediately available. In this article, we address the fundamental question whether any other approach to wave shaping can surpass the limit set by a Fresnel zone plate. In particular, we show for the case of matter waves that the answer to this question is a clear "Yes!", by deriving analytical solutions of the corresponding variational problem. Moreover, we demonstrate the supremacy of our approach compared to Fresnel zone plates by an experiment with light.   

A scalar wave, such as a matter wave or an unpolarized electromagnetic field, comprises two components: an amplitude and a phase. The common way to focus an electromagnetic wave is to modulate its phase using a lens by applying a parabolic phase variation in space. However, there are waves for which a phase-modulating lens does not exist due to technological limitations in implementing phase-altering components in such systems. For instance, implementing such components for X-rays and matter waves often requires subnanometer manufacturing.

More effective approaches to focus waves can be achieved via amplitude modulation in space. For example, blocking part of the wave by a circular aperture or annular rings known as Fresnel zones will focus it to the Arago-Poisson spot \cite{born2013principles}. In these examples, the incoming waves are spatially selected without being modified by the materials. These diffractive focusing techniques are crucially determined by a non-Gaussian initial wave function, as well as by the underlying dimensionality of the problem~\cite{cirone2001quantum,bialynicki2002,case2012diffractive}, and have been employed for surface gravity water waves and plasmonic waves~\cite{Weisman2017,Weisman2021}. 

We emphasize that while Fresnel zones provide one approach to focusing the waves by amplitude modulation, one may question whether other approaches, e.g. nonbinary amplitude modulations, provide even better focusing. In the present article, we obtain the optimal initial wave function for focusing a free particle, i.e. matter waves, in one and two dimensions, and compare and contrast the focusing parameters of the optimal two-dimensional wave function to those of the Fresnel zone approach. The analogy between the Schr\"{o}dinger equation and the paraxial Helmholtz equation allows us to extend our results to electromagnetic waves. Finally, we experimentally verify the focusing properties of the two-dimensional pattern at optical wavelengths using a reflective spatial light modulator and a fabricated transmissive liquid crystal device.\newline

\noindent\textbf{Results} \\
\textbf{Theory of optimal focusing:} 
Our goal is to determine the optimal {\it initial} real-valued, aperture-constrained, and normalized wave function $\psi_0$ in two spatial dimensions that maximizes the intensity $\abs{\psi}^2$ of the field on the symmetry axis at a prescribed focusing time. Our choice of the number of dimensions results from the fact that in one dimension the focusing is weaker, as shown in the Methods section. 

Hence, we assume that $\psi_0$ is radially symmetric, as it provides the best diffractive focusing~\cite{bialynicki2002}, and write the solution as
\begin{equation}\label{Schroedinger-solution}
    \psi(\rho,\tau)=2\pi\int\limits_{0}^{\infty} \rho'{\rm d}\rho'\,{\rm G}^{(2)}(\rho,\tau|\rho',0)\,\psi_0(\rho')
\end{equation}
of the time-dependent two-dimensional Sch\"odinger equation of a free particle in terms of the corresponding Green function 
\begin{equation}\label{Green function}
    {\rm G}^{\rm (2)}(\rho,\tau|\rho',0)=\frac{1}{2\pi {\rm i}\tau} \exp\left({\rm i}\frac{\rho^2+\rho'^2}{2\tau}\right)J_0\left(\frac{\rho\rho'}{\tau}\right)
\end{equation}
with the Bessel function $J_0$ of the first kind~\cite{abramowitz1968handbook}. Here $\rho\equiv r/R$ and $\tau\equiv\hbar t/(MR^2)$ are the dimensionless radial coordinate and time, respectively, wherein $M$ and $R$ denote the mass of the particle and the radius of the circular aperture. In the case of the two-dimensional paraxial Helmholtz equation, $\tau$ is equivalent to the longitudinal distance $z\equiv (kR^2)\tau$ from the screen, where $k$ denotes the wave number. 

We consider only wave functions $\psi_0$ that are truncated by the aperture $\rho\leq 1$ and vanish elsewhere, $\psi_0(\rho \geq 1)=0$. As a result, for a prescribed focusing time $\tau_{\rm f}$, or focal distance $z_{\rm f}\equiv kR^2 \tau_{\rm f}$, the intensity $I[\psi_0]$ along the symmetry axis, $\rho = 0$, takes the form
\begin{eqnarray}\label{intensity_result}
    I[\psi_0]=\frac{1}{\tau_{\rm f}^2}\int\limits_{0}^{1} u {\rm d}u\int\limits_{0}^{1} v {\rm d}v\cos\left(\frac{u^2-v^2}{2\tau_{\rm f}}\right)\psi_0(u)\psi_0(v),\qquad
\end{eqnarray}
where we have used that $\psi_0$ is real.

In order to solve the optimization problem, we first construct the Lagrange function
\begin{equation}\label{Lagrange_function_2D}
    {\mathcal L}[\psi_0]\equiv I[\psi_0]-\lambda\left[2\pi\int\limits_{0}^{1}
    u{\rm d}u\,\psi_0^2(u)-1\right],
\end{equation}
where the Lagrange multiplier $\lambda$ takes into account the normalization condition for $\psi_0$, and then perform the variation of ${\mathcal L}[\psi_0]$ with respect to $\psi_0$, to arrive at the eigenvalue problem
\begin{equation}\label{eigenvalue_problem_2D}
    \frac{1}{2\pi\tau_{\rm f}^2}\int\limits_{0}^{1} v{\rm d}v \cos\left(\frac{u^2-v^2}{2\tau_{\rm f}}\right) \psi_0(v)=\lambda\psi_0(u)
\end{equation}
for the optimal wave function $\psi_0$ corresponding to the eigenvalue $\lambda$. 

Since Eq.~\eqref{eigenvalue_problem_2D} is a linear integral equation with a degenerate kernel, its solution can be found analytically, as shown in the Methods section. Indeed, for a fixed value of $\tau_{\rm f}$, we obtain the maximum eigenvalue
\begin{equation}\label{lambda_max_2D}
\lambda_{+}(\tau_{\rm f})=\frac{1}{8\pi\tau_{\rm f}^2} \left[1+2\tau_{\rm f}\left|\sin{\left(\frac{1}{2\tau_{\rm f}}\right)}\right|\,\right]
\end{equation}
and the normalized optimal initial wave function
\begin{equation}
    \label{eq:optimal}
    \psi_0^{\rm (opt)}(\rho)=N\left[\sqrt{1+a}\cos\left(\frac{\rho^2}{2\tau_{\rm f}}\right)+\sqrt{1-a}\sin\left(\frac{\rho^2}{2\tau_{\rm f}}\right)\right].
\end{equation}
Here, $a\equiv \cos[1/(2\tau_{\rm f})]{\rm sign}\left\{\sin[1/(2\tau_{\rm f})]\right\}$ and $N\equiv 1/\sqrt{8\pi^2\tau_{\rm f}^2\lambda_+(\tau_{\rm f})}$ are the amplitude parameter and the normalization constant, respectively, with ${\rm sign}(x)$ being the sign function. 

Substituting $\psi_0^{\rm (opt)}$ given by Eq.~\eqref{eq:optimal} into the expression, Eq.~\eqref{intensity_result}, for the intensity at $\rho=0$, we prove that the intensity, indeed, achieves its maximum value $I_{\rm max}^{\rm (opt)}(\tau_{\rm f})\equiv I[\psi_0^{\rm (opt)}]=\lambda_+(\tau_{\rm f})$ for any given focusing time $\tau_{\rm f}$, or the dimensionless distance $z_{\rm f}$ from the screen (within the paraxial approximation). In particular, for 
\begin{equation}
 \label{tau_n0}
    \tau_{n_0}\equiv \frac{1}{2\pi n_0},
\end{equation}
where the integer $n_0$ counts the number of Fresnel zones that fit in the circular aperture $0\leq \rho\leq 1$, Eq.~\eqref{lambda_max_2D} yields
\begin{equation} \label{optimal_intensity_result}
   I_{\rm max}^{\rm (opt)}(\tau_{n_0})=\frac{\pi}{2}n_0^2.
\end{equation}

\textbf{Fresnel zones:}
Next we compare the maximum focusing intensity, Eq.~\eqref{optimal_intensity_result}, of the optimal state $\psi_0^{\rm (opt)}$ with the Fresnel zones approach. For this purpose, we consider two different designs.

An amplitude Fresnel zone (AFZ) plate alters the amplitude, while the phase Fresnel zone (PFZ) plate modifies the phase of the incoming wave. In the AFZ, only odd ($n=1,3,5,\ldots$) annular zones are transparent, whereas even ($n=2,4,6,\ldots$) zones are opaque, that is absorbing the incoming waves, with $n=1$ being the innermost zone containing the origin. The AFZ plate is a nonunitary object, i.e. the input intensity is not conserved. In the PFZ, we keep even and odd zones fully transparent; however, the phase in the even zones is shifted by $\pi$. 

\begin{figure*}[htb]
  \includegraphics[width=2.0\columnwidth]{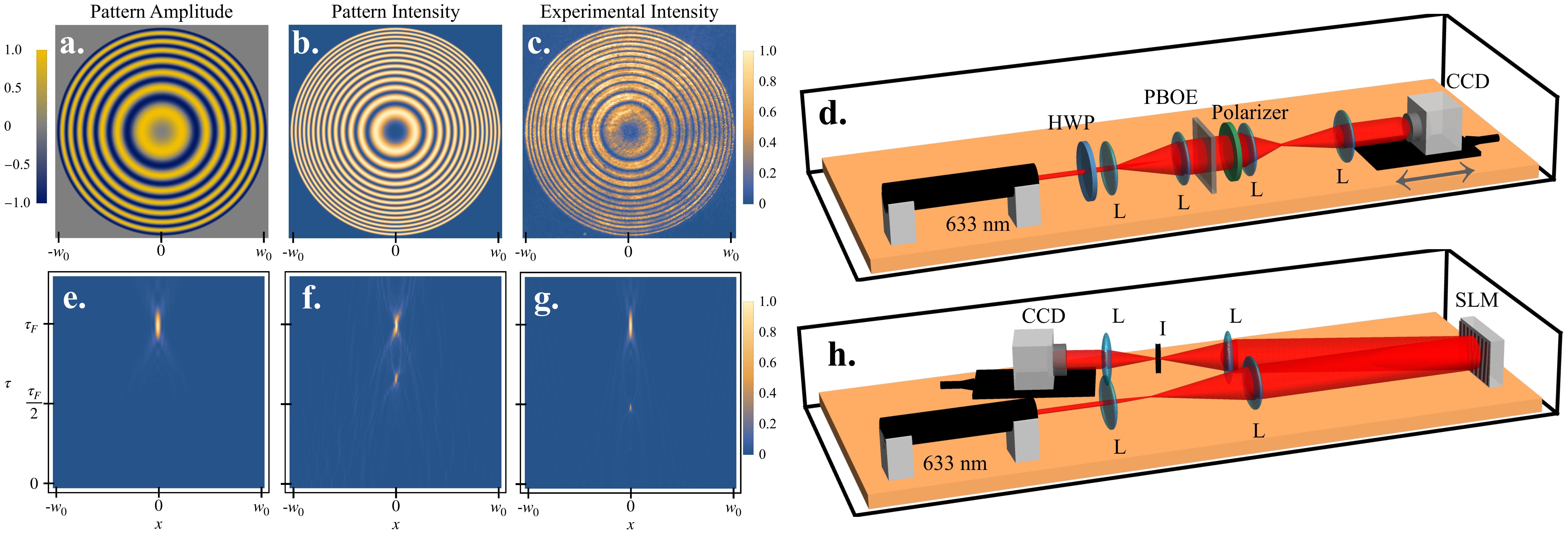}\\
  \caption{{\bf Implementation of Optimal Diffractive Focusing}. We show the theoretical amplitude {\bf (a)} and intensity {\bf (b)} of the $n_0=14$ optimal wave together with the measured optical transmission through the fabricated focusing element {\bf (c)} for the focusing time $\tau_{n_0}=1/(2\pi n_0)$. {\bf (d)} Experimental apparatus used to generate the optimal state with a PBOE. A linearly polarized $633\,{\rm nm}$ Gaussian beam passes a half-wave plate ($\lambda/2$) which rotates it to the horizontal polarization. The beam is then expanded by a factor of five by two lenses (L) to obtain a relatively flat profile before it goes through the PBOE followed by a polarizer. The latter is then imaged by a 4f-system in order to examine its propagation dynamics by a CCD camera. Numerically simulated {\bf (e)} and experimentally observed {\bf (f)} intensity distributions for a cross section of the beam as it propagates from the plane of the device to the focus for the optimal state.  Numerical simulation taking into account contributions {\bf (g)} from both the horizontal (Cosine) and vertical (Sine) polarization components of the modulated beam. Experimental setup {\bf (h)} for focusing with Fresnel zone patterns using the SLM. The 633~nm laser source is expanded to cover the SLM. The 4f-lens system then images the SLM onto the CCD camera with an iris (I) placed at the focus to select the first order of diffraction.}
  \label{Fig1}
\end{figure*}

For the focusing time $\tau_{n_0}$, we derive in the Methods section the maximal intensities
\begin{equation}\label{Fresnel_intensity_1}
    I_{\rm max}^{\rm (AFZ)}(\tau_{n_0})=\frac{2}{\pi}
  \begin{cases}
  n_0 (n_0+1), & \quad n_0=1,3,5,\ldots \\
  n_0^2, & \quad n_0=2,4,6,\ldots
  \end{cases}
\end{equation}
and
\begin{equation}
  \label{Fresnel_intensity_2}
  I_{\rm max}^{\rm (PFZ)}(\tau_{n_0})=\frac{4}{\pi}n_0^2
\end{equation}
at the symmetry axis $\rho=0$. 

A comparison of Eqs. \eqref{optimal_intensity_result}, \eqref{Fresnel_intensity_1} and \eqref{Fresnel_intensity_2} reveals that the optimal state $\psi_0^{\rm (opt)}$ defined by Eq. \eqref{eq:optimal} gives rise to focusing improved by the factor $\pi^2/8$ compared to the best Fresnel method.  

\noindent {\bf Experiments:} Now we demonstrate experimentally optimal diffractive focusing for the two-dimensional case using optical light. For this purpose we have fabricated a transmissive, liquid crystal optical element, that is a Pancharatnam-Berry optical element (PBOE)~\cite{larocque2016arbitrary}, described in the Methods section, which can be operated at many different wavelengths. It generates the optimal state $\psi_0^{\rm (opt)}$ given by Eq.~\eqref{eq:optimal}. The space-varying amplitude for the focusing time $\tau_{n_0}$ defined by Eq.~\eqref{tau_n0} with $n_0=14$ is shown in Fig.~\ref{Fig1}(a) together with the expected and measured intensities, Figs.~\ref{Fig1}(b) and (c), respectively.

The complete experimental apparatus used to generate the optimal state is displayed in Fig.~\ref{Fig1}(d). The PBOE placed between a half-wave plate and a polarizer is illuminated by a 633~nm He-Ne laser with an expanded Gaussian profile. A 4-f lens system is used to image the device on a $1920\times1080$ pixel CCD camera placed on a translation stage, which allows us to measure the intensity of the modulated beam along its propagation to the focus. We have obtained this intensity profile in 50~$\mu\text{m}$ steps for 25.0~mm.

Whereas Fig.~\ref{Fig1}(e) shows the exact evolution of the beam originating from the optimal state $\psi_0^{\rm (opt)}$ given by Eq.~(\ref{eq:optimal}), Figs.~\ref{Fig1}(f) and (g) display the experimentally measured and expected intensity along the propagation. As further elaborated in the Discussion section, we expect imperfections in our optical system to affect the propagation profile.

\begin{figure}[b]
  \includegraphics[width=0.98\columnwidth]{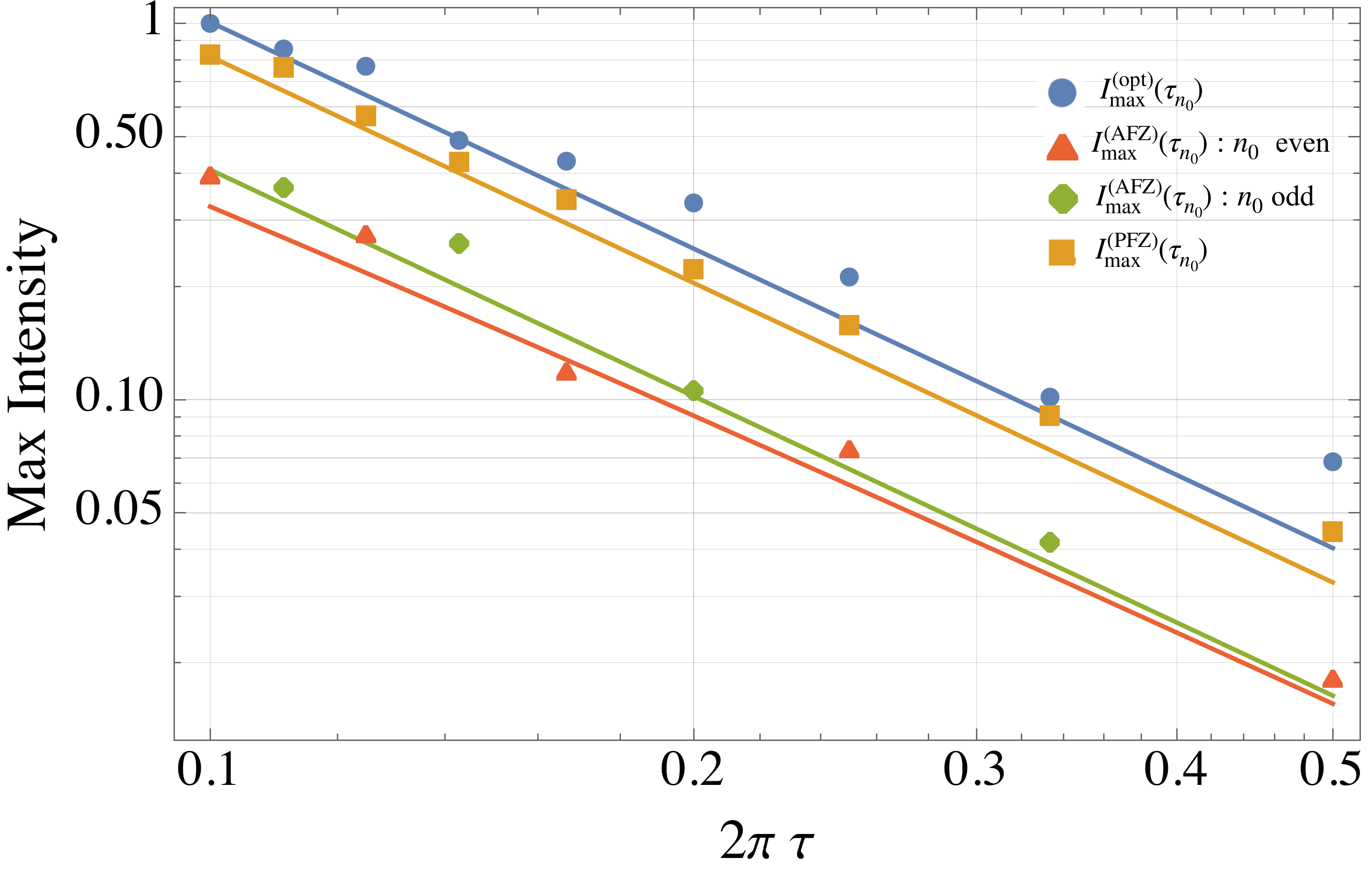}\\
  \caption{
  {\bf Comparison of the three methods for two-dimensional diffractive focusing}. For a fixed focusing time $\tau_{n_0}=1/(2\pi n_0)$, with $n_0=2,3,\ldots,10$, we display the theoretical optimal maximum intensities, $I_{\rm max}^{\rm (opt)}(\tau_{n_0})$ (blue line), given by Eq.~\eqref{optimal_intensity_result}, as well as $I_{\rm max}^{\rm (AFZ)}(\tau_{n_0})$ (red and green), and $I_{\rm max}^{\rm (PFZ)}(\tau_{n_0})$ (orange), associated with the Fresnel zone plates, Eqs.~\eqref{Fresnel_intensity_1} and \eqref{Fresnel_intensity_2}. The points represent the corresponding measurements.}
  \label{Fig2}
\end{figure}

To compare the propagation of $\psi_0^{\rm (opt)}$ with the ones created by the Fresnel zone plates, Eqs.~\eqref{Fresnel_wave_function_1} and \eqref{Fresnel_wave_function_2} in the Methods section, we replace our PBOE by a reflective spatial light modulator (SLM), as depicted in Fig.~\ref{Fig1}(h). 
We used a Hamamatsu liquid crystal on silicon (LCOS) SLM with $1272\times1024$ resolution and a pixel size of $12.5~\mu$m. Moreover, we are able to encode both the intensity and the phase of the pattern on the incident beam using an amplitude masking technique~\cite{bolduc2013exact}. A phase diffraction grating is added to the pattern on the SLM which produces the desired field in the first order of diffraction. We then select this first order with a 4f lens system and an iris, thereby allowing us to remove all other diffraction orders while imaging the SLM plane onto our moveable CCD camera. Although SLMs do not reach the spatial resolution of our PBOE, their programmability can more readily streamline experiments comparing various focusing approaches. 

In Fig.~\ref{Fig2}, we display the maximal focusing intensity $I_\text{max}(\tau_{n_0})$ for nine different patterns corresponding to focusing times $\tau_{n_0}$, as defined in Eq.~\eqref{tau_n0}, with $n_0 = 2,3,\ldots,10$. The results of these experiments involving the optimal wave function as well as both forms of the Fresnel zones are depicted in Fig.~\ref{Fig2}. \newline

\noindent\textbf{Discussion}\\
The propagation of the beam, shown in Fig.~\ref{Fig1}(f), features an artificial peat at $\tau=\tau_{\rm f}/2$ arising from the modulation by our PBOE. Indeed, an imperfect polarization alignment in our generation apparatus leads to contributions of the $\sin (2\alpha)$-term in Eq.~\eqref{q-plate}. The angular orientation of the liquid crystals, $\alpha=\alpha (x,y)$, ranges from 0 to $\pi/2$ such that the horizontally polarized component of the field oscillates from $+1$ to $-1$ according to $\cos(2\alpha)$, while the term $\sin (2\alpha)$ oscillates from 0 to +1 and back to 0. As a consequence, the contribution from the $\sin (2\alpha)$-component, which does not have negative amplitudes, behaves like the Fresnel zones, giving rise to a focus at $\tau_{\rm f}/2$. In Fig.~\ref{Fig1}(g), we show the expected propagation including the contribution from the $\sin (2\alpha)$-term, which is in good agreement with our experimental results shown in Fig.~\ref{Fig1}(f). 

The scalings of the peak intensities of the focusing methods, considered in our article, with $n_0$ are shown in Fig. \ref{Fig2} by solid lines together with the experimental results depicted by the differently coloured data points. The peak focal intensity increases with increasing $n_0$, corresponding to a tighter focusing time $\tau_{n_0}$, or equivalently, to a shorter focal length. Furthermore, the optimal wave function consistently outperforms both methods relying on Fresnel zone plates. 

We conclude this discussion by emphasizing that with our PBOE we were able to achieve a better resolution in our pattern creation, allowing for a tighter focusing time $\tau_{n_0}$ with $n_0=14$, than with the SLM. This advantage is primarily due to the fact that a diffraction grating is necessary when the SLM is used to form an arbitrary wave function, which thus limits the maximum spatial frequency of the phase oscillations corresponding to the desired pattern. In addition, the SLM has limited control over both phase and spatial modulation, as prescribed by its bit depth and pixel pitch, respectively. As $n_0$ is increased, the number of oscillations in $\psi_0^{\rm (opt)}$ from $+1$ to $-1$ increases and in particular, the outer rings of the pattern become ever thinner.

\noindent\textbf{Summary.}
We have derived the optimal real-valued matter wave $\psi_0^{\rm (opt)}$ for focusing in both one- and two- dimensions. The analogy between the Schr\"odinger equation and the paraxial wave equation allows us to transfer our treatment to light. In our optical experiment, we have realized the two-dimensional optimal wave function $\psi_0^{\rm (opt)}$ using liquid crystal devices, verifying the superior focusing properties of $\psi_0^{\rm (opt)}$ compared to diffractive focusing from Fresnel zone patterns. 

The optimal diffractive patterns derived here may be of interest to many different communities where phase modulation, due to technological limitations, is not directly possible. We can also envision extending this technique to vector fields, such as spinors in both optical and matter waves, where combinations of amplitude masks and specially polarized vector modes bring highly structured variations in focused beams~\cite{Dorn:03,wang2008creation,karimi2008improved}. The application of $\psi_0^{\rm (opt)}$ to these tight focusing problems remains to be explored.


\setcounter{equation}{0} \renewcommand{\theequation}{M{\arabic{equation}}}
\footnotesize
\section*{Methods}
\subsection{Optimal state in two dimensions}

To obtain the analytical solution of the integral equation \eqref{eigenvalue_problem_2D}, we cast it in the form
\begin{equation}\label{solution1}
    \psi_0(u)=\frac{1}{2\pi\tau_{\rm f}^2 \lambda}
    \left[
    A\cos\left(\frac{u^2}{2\tau_{\rm f}}\right)+B\sin\left(\frac{u^2}{2\tau_{\rm f}}\right)
    \right],
\end{equation}
where
\begin{equation}
    \label{coefficients-A}
    A\equiv\int\limits_{0}^{1}v\,{\rm d}v \cos\left(\frac{v^2}{2\tau_{\rm f}}\right)\psi_0(v)
\end{equation}
and 
\begin{equation}
    \label{coefficients-B}
    B\equiv\int\limits_{0}^{1}v\,{\rm d}v \sin\left(\frac{v^2}{2\tau_{\rm f}}\right)\psi_0(v)
\end{equation}
are functions solely of $\tau_{\rm f}$.

Next, we insert $\psi_0$ given by Eq. \eqref{solution1} into Eqs. \eqref{coefficients-A} and \eqref{coefficients-B}, and obtain the system 
\begin{align}\label{equation_system_A_B1}
    \left[\lambda-\frac{1+\tau_{\rm f}\sin(1/\tau_{\rm f})}{8\pi\tau_{\rm f}^2}\right]A-\frac{\sin^2[1/(2\tau_{\rm f})]}{4\pi\tau_{\rm f}}B=&0\\ 
    \label{equation_system_A_B2}
    -\frac{\sin^2[1/(2\tau_{\rm f})]}{4\pi\tau_{\rm f}}A+\left[\lambda-\frac{1-\tau_{\rm f}\sin(1/\tau_{\rm f})}{8\pi\tau_{\rm f}^2}\right]B=&0
\end{align}
of algebraic equations for $A$ and $B$, which has non-trivial solutions, only when its determinant is zero, that is
\begin{equation}
    \left(\lambda-\frac{1}{8\pi\tau_{\rm f}^2}\right)^2-\left(\frac{\sin[1/(2\tau_{\rm f})]}{4\pi\tau_{\rm f}}\right)^2=0.
\end{equation}

This elementary quadratic equation has the two solutions 
\begin{equation}
    \lambda_{\pm}=\frac{1}{8\pi\tau_{\rm f}^2}\left[1\pm 2\tau_{\rm f}\left|\sin\left(\frac{1}{2\tau_{\rm f}}\right)\right|\,\right].
\end{equation}

By inserting the maximal eigenvalue $\lambda_{+}$ into Eq. \eqref{equation_system_A_B1}, we find the relation between $A$ and $B$, and thus the normalized optimal initial wave function $\psi_0^{\rm (opt)}$ given by Eq. \eqref{eq:optimal}.

\subsection{Amplitude and phase Fresnel zones: \\ Maximal intensity}

For a given value of $\tau_{\rm f}$ the radii 
\begin{equation} \label{rho_n}
    \rho_n \equiv\sqrt{2\pi\tau_{\rm f} n}
\end{equation}
of the Fresnel zones, with $n=1,2,3,\ldots, n_0$, extend to the maximum number $n_0$ of zones fitting within the circular aperture $0\leq \rho \leq 1$~\cite{born2013principles}. Therefore, the initial wave functions $\psi_0^{\rm (AFZ)}$ and $\psi_0^{\rm (PFZ)}$ for the amplitude and phase Fresnel zone patterns read
\begin{equation}
    \label{Fresnel_wave_function_1}
    \psi_0^{\rm (AFZ)}(\rho)=N_1
    \left[{\mathbb U}(\rho;0,\rho_1)+\sum_{n=1}^{\infty}{\mathbb U}(\rho;\rho_{2n},\rho_{2n+1})\right]
\end{equation}
and
\begin{equation}
    \label{Fresnel_wave_function_2}
   \psi_0^{\rm (PFZ)}(\rho)=\frac{1}{\sqrt{\pi}}
    \left[{\mathbb U}(\rho;0,\rho_1)+\sum_{n=1}^{\infty}(-1)^{n}{\mathbb U}(\rho;\rho_n,\rho_{n+1})\right]
\end{equation}
with ${\mathbb U}(\rho;a,b)\equiv \Theta(b-\rho)-\Theta(a-\rho)$ and $a<b$. Here, $\Theta(\rho)$ denotes the Heaviside function and $N_1$ is a normalization constant depending on $\tau_{\rm f}$. 

To derive an analytical formula for the maximal intensity, we choose the focusing times $\tau_{\rm f}\equiv\tau_{n_0}\equiv 1/(2\pi n_0)$. In this case, Eq. \eqref{rho_n} reduces to $\rho_n=\sqrt{n/n_0}$, and the normalization condition
\begin{equation}
    \pi N_1^2 \left[\rho_1^2+\left(\rho_3^2-\rho_2^2\right)+\left(\rho_5^2-\rho_4^2\right)+\ldots\right]=1    
\end{equation}
for $\psi_0^{\rm (AFZ)}$, Eq. \eqref{Fresnel_wave_function_1}, defines the constant 
\begin{equation}\label{N1}
    N_1\equiv\sqrt{\frac{2}{\pi}}
  \begin{cases}
  \sqrt{\frac{n_0}{n_0+1}}, & \quad n_0=1,3,5,\ldots \\
  1, & \quad n_0=2,4,6,\ldots
  \end{cases}
\end{equation}
as a function of $n_0$. 

Next, we insert the initial profile $\psi_0^{\rm (AFZ)}$ given by Eq. \eqref{Fresnel_wave_function_1} into Eq. \eqref{intensity_result}, and obtain the expression
$$
I_{\rm max}^{\rm (AFZ)}(\tau_{n_0})=\frac{1}{\tau_{n_0}^2}\left|\int\limits_{0}^{1}u{\rm d}u\,\exp\left({\rm i}\frac{u^2}{2\tau_{n_0}}\right)\psi_0^{\rm (AFZ)}(u)\right|^2
$$
$$
=N_1^2\left|\exp\left({\rm i}\frac{\rho_1^2}{2\tau_{n_0}}\right)-1+\exp\left({\rm i}\frac{\rho_3^2}{2\tau_{n_0}}\right)-\exp\left({\rm i}\frac{\rho_2^2}{2\tau_{n_0}}\right)+\ldots\right|^2,
$$
that is
\begin{equation}
 \label{I_AFZ_intensity}
  I_{\rm max}^{\rm (AFZ)}(\tau_{n_0})=N_1^2
  \begin{cases}
  (n_0+1)^2, & \quad n_0=1,3,5,\ldots \\
  n_0^2, & \quad n_0=2,4,6,\ldots,
  \end{cases}
\end{equation}
where we have used the fact that $\rho_n^2/(2\tau_{n_0})=n\pi$. 

As a result, Eq. \eqref{I_AFZ_intensity} combined with Eq. \eqref{N1} gives rise to the maximum intensity $I_{\rm max}^{\rm (AFZ)}(\tau_{n_0})$, Eq. \eqref{Fresnel_intensity_1}, produced by the amplitude Fresnel zones. Analogously, we derive the corresponding maximum intensity $I_{\rm max}^{\rm (PFZ)}(\tau_{n_0})$, Eq. \eqref{Fresnel_intensity_2}, for the Fresnel phase zones.

\subsection{Optimal state in one dimension}
In this section we determine the optimal initial real-valued and normalized wave function $\varphi_0\equiv\varphi_0(x)$ that maximizes the intensity $\abs{\varphi(0)}^2$ of the field at $x=0$, at the focusing time $t_{\rm f}$.

In this case we use the one-dimensional Green function 
\begin{equation}\label{Green function_1d}
    {\rm G}^{\rm (1)}(\xi,\tau|\xi',0)=\frac{1}{\sqrt{2\pi {\rm i}\tau}} \exp\left[{\rm i}\frac{(\xi-\xi')^2}{2\tau}\right]
\end{equation}
for the time-dependent one-dimensional Schr\"odinger equation of a free particle. Here, $\xi\equiv x/L$ and $\tau\equiv\hbar t/(ML^2)$ are the dimensionless position and time, respectively, and $M$ and $L$ denote the mass of the particle and the slit width.

We again apply the method of the Lagrange multipliers and arrive at the eigenvalue problem
\begin{equation}\label{eigenvalue_problem_1d}
    \frac{1}{2\pi\tau_{\rm f}}\int\limits_{-1}^{1} {\rm d}\xi'\cos\left(\frac{\xi^2-\xi'^2}{2\tau_{\rm f}}\right)\varphi_0(\xi')=\mu\varphi_0(\xi)
\end{equation}
for the optimal initial wave function $\varphi_0$ with the eigenvalue $\mu$, that determines the maximum intensity achieved at $\tau_{\rm f}$. Here, we have assumed that $\varphi_0(\xi)=0$ for $|\xi|>1$.   

Since we are interested in the maximum of the intensity, we solve the integral equation~\eqref{eigenvalue_problem_1d} only for the largest eigenvalue $\mu_+(\tau_{\rm f})$. As a result, for a given $\tau_{\rm f}$, we find the optimal initial wave function
\begin{equation}\label{psi_max_1d}
    \varphi_0^{\rm (opt)}(\xi)=\sqrt{\frac{1+r}{4\pi\tau_{\rm f}\mu_+}}\cos\left(\frac{\xi^2}{2\tau_{\rm f}}\right)+\sqrt{\frac{1-r}{4\pi\tau_{\rm f}\mu_+}}\sin\left(\frac{\xi^2}{2\tau_{\rm f}}\right)
\end{equation}
for $|\xi|\leq 1$, with $\varphi_0^{\rm (opt)}(\xi)=0$ for $|\xi|>1$, and the corresponding maximal eigenvalue
\begin{equation}\label{lambda_max_1D}
    \mu_+(\tau_{\rm f})=I_{\rm max}^{\rm (1D)}(\tau_{\rm f})=\mathcal{I}\left[\sqrt{2/(\pi\tau_{\rm f})}\right].
\end{equation}
Here we have expressed the intensity 
$$
\mathcal{I}(z)\equiv\frac{1}{4}\left\{z^2+z\sqrt{\left[C\left(z\right)\right]^2+\left[S\left(z\right)\right]^2}\right\}
$$
and the parameter
\begin{equation}\label{parameter_r}
    r(\tau_{\rm f})\equiv\frac{C\left[\sqrt{2/(\pi\tau_f)}\,\right]}{\sqrt{\left\{C\left[\sqrt{2/(\pi\tau_f)}\,\right]\right\}^2+\left\{S\left[\sqrt{2/(\pi\tau_f)}\,\right]\right\}^2}}
\end{equation}
in terms of the Fresnel integrals $C$ and $S$ \cite{abramowitz1968handbook}.

\subsection{Pancharatnam-Berry optical element}

Our device consists of a patterned layer of birefringent nematic liquid crystals whose orientation locally determines that of the medium's optical axis.  This feature causes the element to have the action 
\begin{eqnarray}\label{q-plate}
  \hat{\mathcal U}_q & \cdot & \begin{pmatrix}\mathbf{e}_{\rm H} \\ \mathbf{e}_{\rm V}\end{pmatrix} = 
  \cos\left(\frac{\delta}{2}\right)\begin{pmatrix}\mathbf{e}_{\rm H} \\ \mathbf{e}_{\rm V}\end{pmatrix} \nonumber \\ 
  &+& i\sin\left(\frac{\delta}{2}\right)
  \begin{pmatrix}
  \cos\left[2\alpha(x,y)\right]\;\;\;\;\; \sin\left[2\alpha(x,y)\right] \\ 
  \sin\left[2\alpha(x,y)\right]\; -\cos\left[2\alpha(x,y)\right]
  \end{pmatrix} 
  \begin{pmatrix}\mathbf{e}_{\rm H} \\ \mathbf{e}_{\rm V}\end{pmatrix}\nonumber \\
\end{eqnarray}
on the horizontal ${\bf e}_{\rm H}$ and vertical ${\bf e}_{\rm V}$ polarization components of an optical beam. Here $\delta$ is the optical retardation of the liquid crystal molecules and $\alpha\equiv\alpha(x,y)$ is the device's spatially dependent liquid crystal axis orientation expressed in terms of the transverse Cartesian coordinates $x$ and $y$.

When the device is perfectly tuned, that is for $\delta=\pi$, and followed by a {\it horizontally oriented polarizer}, it can effectively be used to mask the amplitude profile of incoming {\it horizontally polarized light} by a factor of $\cos\left[2\alpha(x,y)\right]$. This procedure was employed to generate our real-valued optimal state $\psi_0^{\rm (opt)}\equiv\psi_0^{\rm (opt)}(x,y)$ by means of a device defined by an optical axis of $\alpha(x,y)=(1/2)\arccos\left[\psi_0^{\rm (opt)}(x,y)/\psi_{\rm max}^{\rm (opt)}\right]$, where  $\psi_{\rm max}^{\rm (opt)}$ is the maximum value of the optimal state.

\bibliographystyle{naturemag}
\bibliography{OptimalFocusing}

\providecommand{\noopsort}[1]{}
\begin{thebibliography}{10}
\expandafter\ifx\csname url\endcsname\relax
  \def\url#1{\texttt{#1}}\fi
\expandafter\ifx\csname urlprefix\endcsname\relax\def\urlprefix{URL }\fi
\providecommand{\bibinfo}[2]{#2}
\providecommand{\eprint}[2][]{\url{#2}}

\bibitem{born2013principles}
\bibinfo{author}{Born, M.} \& \bibinfo{author}{Wolf, E.}
\newblock \emph{\bibinfo{title}{Principles of Optics: Electromagnetic Theory of
  Propagation, Interference and Diffraction of Light}}
  (\bibinfo{publisher}{Elsevier}, \bibinfo{year}{2013}).

\bibitem{cirone2001quantum}
\bibinfo{author}{Cirone, M.~A.}, \bibinfo{author}{Rza\ifmmode \mbox{\c{}}\else
  \c{}\fi{}\ifmmode~\dot{z}\else \.{z}\fi{}ewski, K.},
  \bibinfo{author}{Schleich, W.~P.}, \bibinfo{author}{Straub, F.} \&
  \bibinfo{author}{Wheeler, J.~A.}
\newblock \bibinfo{title}{Quantum anticentrifugal force}.
\newblock \emph{\bibinfo{journal}{Phys. Rev. A}} \textbf{\bibinfo{volume}{65}},
  \bibinfo{pages}{022101} (\bibinfo{year}{2001}).

\bibitem{bialynicki2002}
\bibinfo{author}{Bia\l{}ynicki-Birula, I.}, \bibinfo{author}{Cirone, M.~A.},
  \bibinfo{author}{Dahl, J.~P.}, \bibinfo{author}{Fedorov, M.} \&
  \bibinfo{author}{Schleich, W.~P.}
\newblock \bibinfo{title}{In- and outbound spreading of a free-particle
  $s$-wave}.
\newblock \emph{\bibinfo{journal}{Phys. Rev. Lett.}}
  \textbf{\bibinfo{volume}{89}}, \bibinfo{pages}{060404}
  (\bibinfo{year}{2002}).

\bibitem{case2012diffractive}
\bibinfo{author}{Case, W.~B.}, \bibinfo{author}{Sadurni, E.} \&
  \bibinfo{author}{Schleich, W.~P.}
\newblock \bibinfo{title}{A diffractive mechanism of focusing}.
\newblock \emph{\bibinfo{journal}{Opt. Express}} \textbf{\bibinfo{volume}{20}},
  \bibinfo{pages}{27253} (\bibinfo{year}{2012}).

\bibitem{Weisman2017}
\bibinfo{author}{Weisman, D.} \emph{et~al.}
\newblock \bibinfo{title}{Diffractive focusing of waves in time and in space}.
\newblock \emph{\bibinfo{journal}{Phys. Rev. Lett.}}
  \textbf{\bibinfo{volume}{118}}, \bibinfo{pages}{154301}
  (\bibinfo{year}{2017}).

\bibitem{Weisman2021}
\bibinfo{author}{Weisman, D.} \emph{et~al.}
\newblock \bibinfo{title}{Diffractive guiding of waves by a periodic array of
  slits}.
\newblock \emph{\bibinfo{journal}{Phys. Rev. Lett.}}
  \textbf{\bibinfo{volume}{127}}, \bibinfo{pages}{014303}
  (\bibinfo{year}{2021}).

\bibitem{abramowitz1968handbook}
\bibinfo{author}{Abramowitz, M.} \& \bibinfo{author}{Stegun, I.~A.}
\newblock \emph{\bibinfo{title}{Handbook of Mathematical Functions with
  Formulas, Graphs, and Mathematical Tables}}, vol.~\bibinfo{volume}{55}
  (\bibinfo{publisher}{US Government Printing Office}, \bibinfo{year}{1968}).

\bibitem{larocque2016arbitrary}
\bibinfo{author}{Larocque, H.} \emph{et~al.}
\newblock \bibinfo{title}{Arbitrary optical wavefront shaping via spin-to-orbit
  coupling}.
\newblock \emph{\bibinfo{journal}{Journal of Optics}}
  \textbf{\bibinfo{volume}{18}}, \bibinfo{pages}{124002}
  (\bibinfo{year}{2016}).

\bibitem{bolduc2013exact}
\bibinfo{author}{Bolduc, E.}, \bibinfo{author}{Bent, N.},
  \bibinfo{author}{Santamato, E.}, \bibinfo{author}{Karimi, E.} \&
  \bibinfo{author}{Boyd, R.~W.}
\newblock \bibinfo{title}{Exact solution to simultaneous intensity and phase
  encryption with a single phase-only hologram}.
\newblock \emph{\bibinfo{journal}{Opt. Lett.}} \textbf{\bibinfo{volume}{38}},
  \bibinfo{pages}{3546--3549} (\bibinfo{year}{2013}).

\bibitem{Dorn:03}
\bibinfo{author}{Dorn, R.}, \bibinfo{author}{Quabis, S.} \&
  \bibinfo{author}{Leuchs, G.}
\newblock \bibinfo{title}{Sharper focus for a radially polarized light beam}.
\newblock \emph{\bibinfo{journal}{Phys. Rev. Lett.}}
  \textbf{\bibinfo{volume}{91}}, \bibinfo{pages}{233901}
  (\bibinfo{year}{2003}).

\bibitem{wang2008creation}
\bibinfo{author}{Wang, H.}, \bibinfo{author}{Shi, L.},
  \bibinfo{author}{Lukyanchuk, B.}, \bibinfo{author}{Sheppard, C.} \&
  \bibinfo{author}{Chong, C.~T.}
\newblock \bibinfo{title}{Creation of a needle of longitudinally polarized
  light in vacuum using binary optics}.
\newblock \emph{\bibinfo{journal}{Nature Photonics}}
  \textbf{\bibinfo{volume}{2}}, \bibinfo{pages}{501--505}
  (\bibinfo{year}{2008}).

\bibitem{karimi2008improved}
\bibinfo{author}{Karimi, E.}, \bibinfo{author}{Piccirillo, B.},
  \bibinfo{author}{Marrucci, L.} \& \bibinfo{author}{Santamato, E.}
\newblock \bibinfo{title}{Improved focusing with hypergeometric-gaussian
  type-ii optical modes}.
\newblock \emph{\bibinfo{journal}{Opt. Express}} \textbf{\bibinfo{volume}{16}},
  \bibinfo{pages}{21069} (\bibinfo{year}{2008}).

\end{thebibliography}

\vspace{2 EM}
\noindent\textbf{Data availability}
\noindent
The data that support the findings of this study are available from the corresponding author upon reasonable request.
\vspace{1 EM}

\noindent\textbf{Code availability}
\noindent
The code used for the data analysis is available from the corresponding author upon reasonable request.

\vspace{1 EM}
\noindent\textbf{Ethics declarations} The authors declare no competing interests.

\vspace{1 EM}
\noindent\textbf{Corresponding authors}
Correspondence and requests for materials should be addressed to maxim.efremov@dlr.de.

\vspace{1 EM}
\noindent\textbf{Acknowledgement}
Maxim A. Efremov and Felix Hufnagel contributed equally to this work. We thank P. Boegel for fruitful discussions. W.P.S. is grateful to the Hagler Institute for Advanced Study at Texas A$\&$M University for a Faculty Fellowship, and to Texas A$\&$M AgriLife Research for the support of this work. This project was conceived during a visit of E.K. to Ulm University made possible by {\it IQST}. F. H. and  E.K. acknowledge the support of the Canada Research Chair (CRC) Program, NRC-uOttawa Joint Centre for Extreme Quantum Photonics (JCEP) and NSERC.

\newpage

\end{document}